\documentclass[12pt]{article}

\usepackage[pdftex]{graphicx}
\usepackage{cite}

\usepackage{multirow}%
\usepackage{amsmath,amssymb,amsfonts}%
\usepackage{amsthm}%
\usepackage{mathrsfs}%
\usepackage{xcolor}%
\usepackage{authblk}
\usepackage{hyperref}
\usepackage{microtype}

\begin{document}

\title{Description of Bhabha-scattering at small angles using the Monte Carlo generator {\tt ReneSANCe}}

\author[1]{A.~Arbuzov}
\author[1]{S.~Bondarenko}
\author[1]{I.~Boyko}
\author[1,2]{Ya.~Dydyshka}
\author[1]{L.~Kalinovskaya}
\author[1]{A.~Kampf}
\author[1]{R.~Sadykov}
\author[1]{A.~Tropina }
\author[1,2]{V.~Yermolchyk}
\author[1,2]{Yu.~Yermolchyk}

\affil[1]{\small Joint Institute for Nuclear Research, Dubna, 141980}
\affil[2]{\small Research Institute for Nuclear Problems of Belarusian State University, Minsk, 220030}

\date{}

\maketitle

\abstract{
Results of a simulation of Bhabha small-angle scattering process at center-of-mass energy of 240~GeV and at the $Z$-boson resonance using the Monte Carlo generator {\tt{ReneSANCe}} are presented. For the given accuracy about $10^{-4}$ in estimating luminosity of future electron-positron accelerators, the minimum cutoff angle for simulated events is established.
}

\vspace*{6pt}


\label{sec:intro}
\section*{Introduction}

The Bhabha scattering process is a background to numerous two-photon production processes and a key component in determining the beam lifetime. It is one of the three basic processes for estimating the luminosity of electron-positron colliders. There are several Monte-Carlo generators which help to estimate Bhabha scattering cross section: \href{https://www2.pv.infn.it/~hepcomplex/babayaga.html} {{\tt{BabaYaga}}} \cite{CarloniCalame:2003yt,BALOSSINI2006227,BALOSSINI2008209,CARLONICALAME200116,CarloniCalame:2000pz}, \href{https://annapurna.ifj.edu.pl/programs/programs.html}{{\tt{BHLUMI}}} \cite{Jadach:1996is}, \href{https://placzek.web.cern.ch/placzek/}{{\tt BHWIDE}} \cite{Jadach:1995nk}, {\tt MCGPJ} \cite{Arbuzov:2005pt}, \href{https://renesance.hepforge.org/}{{\tt{ReneSANCe}}} \cite{Sadykov:2020any}. Only the first two of them are widely used to estimate luminosity at small scattering angles ({\tt SABS} --- Small Angle Bhabha Scattering).

The description of the process 
\begin{eqnarray}
\label{bhabha}
e^{+}(p_1) + e^{-}(p_2) \rightarrow 
e^-(p_3) + e^+(p_4) (+ \gamma (p_5))
\end{eqnarray}
at the electroweak one-loop level and the results of analytical calculations of scalar form factors and helicity amplitudes of polarized Bhabha scattering in the {\tt SANC} system were presented in \cite{Bardin:2017mdd}. Implementation details  of this process in the Monte-Carlo generator {\tt ReneSANCe} were described in ~\cite{Sadykov:2020any}.

The purpose of this work is to investigate Bhabha scattering at arbitrarily small or even vanishingly small scattering angles.
As will be shown below, at such angles additional potentially significant contributions to the measured value of the {\tt SABS} cross section arise.

\section{Numerical results} 

The numerical results obtained by {\tt ReneSANCe} and {\tt BHLUMI} are in good agreement with each other within the limits of systematic errors. Numerical calculations were performed with allowance for the conditions of the CERN workshop "Event generators for Bhabha scattering" \footnote{\footnotesize{Jadach, S. and others, {\it Event generators for Bhabha scattering}, CERN Workshop on LEP2 Physics (followed by 2nd meeting, 15-16 Jun 1995 and 3rd meeting 2-3 Nov 1995), hep-ph/9602393}}: the one-loop level approximation without taking into account the $Z$-exchange contribution, up-down interference, vacuum polarization. The results are presented at different values of the energy cut $z_{min} = s'/s$, where $\sqrt{s'}$ is the energy of collision without taking into account initial state radiation.

For instance, in the case of calorimetric event selection at $z_{min} = 0.1$ the {\tt ReneSANCe} cross section is $\sigma,\text{pb} = 166.33(1)$ and the {\tt BHLUMI} cross section is $\sigma,\text{pb} = 166.33(2)$; at 
$z_{min} = 0.7$ $\sigma,\text{pb} = 161.77(1)$ and $\sigma,\text{pb} = 161.79(2)$, respectively.

To conduct the research, 100 million events of the process in question were generated using {\tt ReneSANCe}. The events were analyzed at two center-of-mass (c.m.s.) energies $\sqrt{s}=$ 91.18 GeV and 240 GeV, considering that each arm of the luminometer detected an electron (positron) and/or photon(s).

The results were obtained in the electroweak $\alpha(0)$ scheme using a set of parameters:
\begin{table}[ht!]
    \centering
    \begin{tabular}{lcllclc}
    $\alpha^{-1}(0)$ &=& 137.035999084    &            & &                & \\
    $M_W$            &=& 80.379 GeV       & $\Gamma_W$ &=& 2.0836 GeV     & \\
    $M_Z$            &=& 91.1876 GeV      & $\Gamma_Z$ &=& 2.4952 GeV     & \\
    $M_H$            &=& 125.0 GeV        & $m_e$      &=& 0.51099895 MeV & \\
    $m_\mu$          &=& 0.1056583745 GeV & $m_\tau$   &=& 1.77686 GeV    & \\
    $m_d$            &=& 0.083 GeV        & $m_s$      &=& 0.215 GeV      & \\
    $m_b$            &=& 4.7 GeV          & $m_u$      &=& 0.062 GeV      & \\
    $m_c$            &=& 1.5 GeV          & $m_t$      &=& 172.76 GeV.    &
    \end{tabular}
    \label{parameters}
\end{table}

In addition, the following kinematic cuts were also taken into account:

\noindent

 $\bullet$ The scattering angle of electrons is any value, up to zero;
  \noindent

 $\bullet$ Angular acceptance of the luminometer is $ 30 ~\mbox{\text{mrad}} < \theta < 174.5~\mbox{\text{mrad}}.$

We considered two types of event selection (ES): (a)~{\bf ES-BARE} (non-calorimetric) and (b)~{\bf ES-CALO} (calorimetric).
In case (a), events are selected when each arm of the luminometer was excited by an electron (positron) with the energy $E_{e^{\pm}} > 0.5E_{\text{beam}}$, photons ignored. 
In case (b), events are selected, when one of the luminometer arms was excited by an electron with the energy $E_{e^{\pm}} > 0.5E_{\text{beam}}$, and the other arm was excited by either an electron (positron) or/and a photon(s) with the total energy $E_{e,\gamma} > 0.5E_{\text{beam} }$.

The one-loop cross section of the process under consideration has the form
\begin{eqnarray}
\label{sigma1}
\sigma^{\mathrm{one-loop}} = 
  \sigma^{\mathrm{Born}}
+ \sigma^{\mathrm{QED}} 
+ \sigma^{\mathrm{weak}}.
\end{eqnarray}
The Born level cross section $\sigma^{\rm Born}$ was calculated with allowance for $Z$ and $\gamma$ exchange.
The gauge-invariant QED contribution $\sigma^{\mathrm{QED}}$ was estimated independently. The electroweak contribution $\sigma^{\rm weak}$ was defined as the difference between the full one-loop electroweak correction and its QED part.

The results of calculating various contributions of radiation corrections are presented in the form of relative corrections
$$\delta^{\text{contr.}} = \dfrac{\sigma^{\text{contr.}}}{\sigma^{\text{Born}}}.$$

\underline{\bf Analysis of {\bf ES-BARE}\&{\bf ES-CALO} event selections}

The results of calculating various contributions of radiative corrections for the event selection of type (a) and the cross section at the Born level for both energies are given in Table \ref{ResultsBARE1}.  
\begin{table}[h]
	\centering
		\begin{tabular}{lcc}
			\hline\hline
			$\sqrt{s}$, GeV & $91.18$ & $240$ \\
			\hline
   
			$\sigma^{\rm Born}$, pb & $135008.457(1)$ & $19473.628(2)$ \\

			$\delta^{\rm one-loop}$, \% & $-1.509(2)$ & $-0.828(3)$\\
            \hline
			$\delta^{\rm QED}$, \% &  $-6.170(2)$ & $-7.002(3)$\\

			$\delta^{\rm weak}$, \% & $0.00918(3)$ & $-0.00675(2)$\\
   
			$\delta^{\rm VP}$, \% & $4.65262$ & $6.18667$ \\

    		$\delta^{\rm weak+VP}$, \% & $4.66180(3)$ & $6.17992(3)$ \\
            \hline
            \hline
		\end{tabular}
        \caption{Calculation results for various contributions of radiative corrections to the total cross section at the c.m.s. energies $\sqrt{s}$ = 91.18 GeV and 240 GeV for type (a) event selection.}
        \label{ResultsBARE1}
\end{table}

The results of calculating the QED contribution for event selections of type (a) and (b) are given in Table \ref{ResultsBARECALO}.
The Bhabha scattering cross sections obtained for type (b) events are $3\%$ larger than the cross sections obtained for type (a) events at both c.m.s. energies. Most of the difference is due to events with a collinear photon or events where the electron is scattered at an angle larger than the luminometer's acceptance angle  while a hard photon emitted from the initial state enters the luminometer.

In addition, we observe a very important effect
\footnote{The effect is estimated as the difference between the last two rows of Table \ref{ResultsBARECALO}.}:
approximately ${\sim 0.14}$ \% of the total cross section
is accounted for by events with electron scattering angles below the acceptance of the luminometer.
It would be a serious mistake to miss such events from the analysis.
\begin{table}[!h]
\begin{center}
 \begin{tabular}{lcc}
	\hline\hline
	$\sqrt{s}$, GeV & $91.18$ & $240$\\ 
    \hline
    $\delta^{\rm QED}$({\rm ES-BARE}), \% &  $-6.170(2)$    & $-7.002(3)$     \\
    $\delta^{\rm QED}$({\rm ES-CALO}), \% &  $-3.486(4)$    & $-3.992(4)$      \\ 
   $\delta^{\rm QED}({\rm \vartheta > 0.030\; \text{rad}})$, \% &  $-3.349(2)$           & $-3.854(4)$\\
   \hline
   $\delta^{\rm QED}({\rm \vartheta < 0.030\; \text{rad}}), \%$   & \;\;\;$0.137(4)$& \;\;\;$0.138(6)$\\
   \hline \hline
	\end{tabular}
\end{center}
 \caption{
 QED relative corrections for event selections of type (a) and (b) for the c.m.s. energies  $\sqrt{s}$ = 91.18 and 240 GeV.}
   \label{ResultsBARECALO} 
\end{table}

\underline{\bf Analysis of angular distributions}

Let us illustrate the obtained effect by the example of distributions of the fraction of the number of events over lepton scattering angle $\vartheta_{14}$, i.e., the angle between the particle with momentum $p_1$ (initial positron) and the particle with momentum $p_4$ (final positron).
\begin{figure}[!h]
\begin{center}
\includegraphics[width=0.35\textwidth]{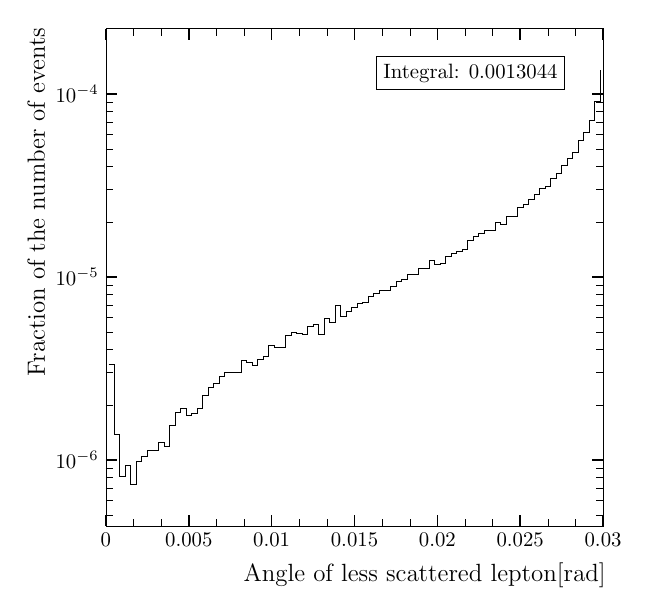}
\vspace{-3mm}
\includegraphics[width=0.35\textwidth]{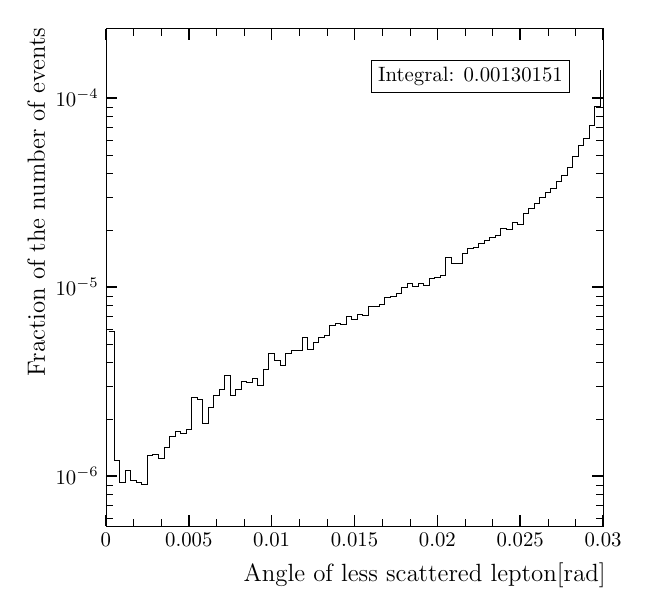}
\caption{Angular distribution in the range $(0;0.03)$ rad for the c.m.s. energies $\sqrt{s} = 91.18$ GeV (left) and $240$ GeV (right).}
\label{fig01}
\end{center}
\vspace{-5mm}
\end{figure}

\begin{figure}[!h]
\begin{center}
        \includegraphics[width=0.35\textwidth]{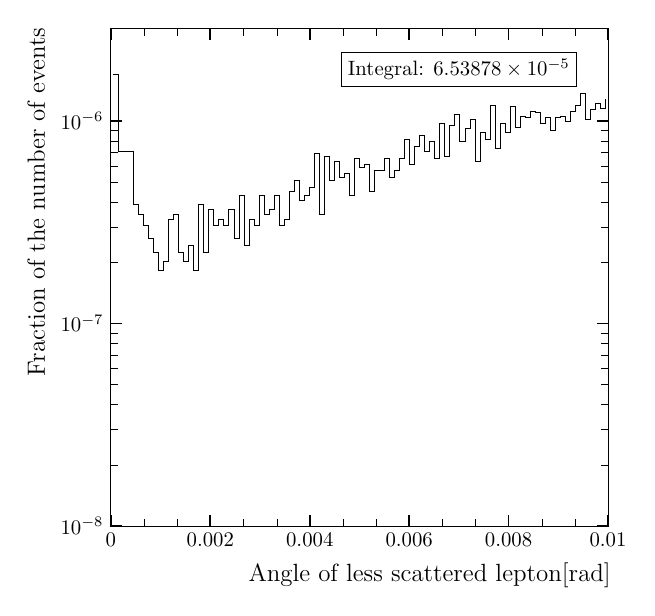} 
        \includegraphics[width=0.35\textwidth]{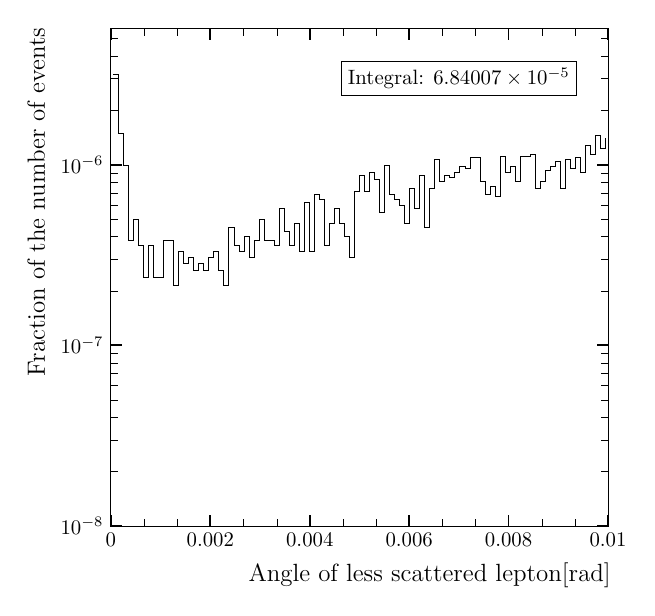}
    \caption{Angular distribution in the range $(0;0.01)$ rad for the c.m.s. energies $\sqrt{s} = 91.18$ GeV (left) and $240$ GeV (right).}
    \label{fig02}
 \end{center}
\end{figure}
In Fig. \ref{fig01} we see a significant dependence of the Bhabha scattering cross section on the minimum scattering angle. The contribution to the cross section from events at a scattering angle in the range (0;30) mrad is approximately $0.13$~\%. 
With the required luminosity estimation accuracy of $10^{-4}$~\cite{FCC:2018byv,CEPCStudyGroup:2018ghi}, it is necessary to take into account events from 10 to 30~mrad  (Fig.~\ref{fig01}), and events with  emission angle of an electron less than 10~mrad (Fig.~\ref{fig02}) can be neglected, since their contribution corresponds to approximately $0.0065$~\% of the total cross section. 

\section{Conclusion}

The Bhabha small-angle scattering is one of the traditional processes that can be used to estimate the luminosity of electron-positron colliders. 
We investigated this process using the Monte Carlo generator {\tt ReneSANCe}.
A technical comparison showed good agreement  
with the results obtained with the Monte Carlo generator {\tt BHLUMI}.
Due to the unique ability of the the {\tt ReneSANCe} generator to operate in the full kinematic phase space,
events were analyzed at various electron scattering angles down to zero.
Our studies of a sample of events in the range from 10 to 30 mrad show that neglecting them leads to a luminosity shift of $0.13 - 0.14$~\% both at the $Z$-resonance and at the c.m.s. energy 240~GeV. The contribution to the cross section from events with smaller scattering angles can be neglected, since their order does not exceed $\sim 10^{-5}$.

In the future, it is planned to improve the {\tt ReneSANCe} generator by taking into account higher-order radiation corrections and exponential photon radiation.

\section{Funding}
\label{sec:funding}
The research is supported by the Russian Science Foundation (project No. 22-12-00021).

\end{document}